\documentclass[a4paper,fleqn,usenatbib]{mnras}

\usepackage{newtxtext,newtxmath}

\usepackage[T1]{fontenc}
\usepackage{ae,aecompl}

\usepackage{graphicx}	% Including figure files
\usepackage{amsmath}	% Advanced maths commands
\usepackage{amssymb}	% Extra maths symbols

\title[correcting the period derivatives]{Correcting the measured values of the rate of change of the spin and orbital periods of rotation powered pulsars}

\author[Pathak \& Bagchi]{
Dhruv Pathak,$^{1,2}$\thanks{E-mail: dhruvpathak@imsc.res.in (DP)}
Manjari Bagchi,$^{1,2}$
\\
$^{1}$The Institute of Mathematical Sciences, C. I. T. campus, Taramani, Chennai, 600113, India\\
$^{2}$Homi Bhabha National Institute, Training School Complex, Anushakti Nagar, Mumbai 400094, India
}

\date{Accepted XXX. Received YYY; in original form ZZZ}

\pubyear{2019}

\begin{document}
\label{firstpage}
\pagerange{\pageref{firstpage}--\pageref{lastpage}}
\maketitle

% Abstract of the paper
\begin{abstract}
For rotation powered pulsars, the rate of change of the spin period is expected to be positive. On the other hand, for a clean binary system (where gravity is the only force acting on the pulsar) the rate of change of the orbital period is expected to be negative. There are, however, some pulsars with the measurements of the rate of change of the spin period as negative quantities. Likewise, there are some pulsars with positive measurements of the rate of change of the orbital period. We investigate these cases by eliminating the external dynamical effects. We also look at those cases where at first, the measured values possess the correct sign, but on subtracting the dynamical effects, the sign changes. We use `GalDynPsr' package to perform these tasks. Moreover, we investigate possible reasons for such anomalies. 
\end{abstract}


%%%%%%%%%%%%%%%%%%%%%%%%%%%%%%%%%%%%%%%%%%%%%%%%%%

%%%%%%%%%%%%%%%%% BODY OF PAPER %%%%%%%%%%%%%%%%%%

\section{Introduction}

The spin period, as well as, the orbital period (in case of a binary system) of a pulsar are among the quantities that vary with time. The mis-alignment of the spin axis and the magnetic axis of a pulsar leads to a time-varying magnetic moment, which, in turn, is responsible for the emission of the electromagnetic radiation. This electromagnetic power is accounted for by the loss of the rotational kinetic energy of the pulsar. As a result, the pulsar slows down, i.e., its spin period increases. This leads to the rate of change of the spin period being a positive quantity. On the other hand, in case of a binary system, where gravity is the only interaction between the pulsar and its companion, the quadrupolar gravitational radiation leads to the orbital shrinkage and consequent increase in the angular orbital frequency (due to angular momentum conservation). Hence, the orbital period decreases with time, making the rate of change of the orbital period a negative quantity. 

However, the above mentioned effects are intrinsic to the pulsar system. The observed values of the rates of change of periods, on the other hand, are affected by external factors like the line of sight relative velocity and acceleration of the pulsar with respect to the solar system barycentre. The acceleration of the pulsar can be affected by various effects like the gravitational potential of the Galaxy, the cluster potential (if the pulsar resides in a Globular cluster), the gravitational force by near-by stars, gravitational force due to the presence of near-by dense molecular clouds, etc. These effects have been discussed in detail in \citet{pb18}. These effects can cause, in some cases, the observed values to posses a sign opposite to the expected sign of the intrinsic rate of change of periods. The intrinsic rate of change of orbital period is given as:

\begin{equation}
\dot{P}_{\rm b, int} = \dot{P}_{\rm b, obs} - \dot{P}_{\rm b, Gal} - \dot{P}_{\rm b, Shk} 
\label{eq:pbint1}
\end{equation} 

where $\dot{P}_{\rm b, obs}$ is the observed rate of the change of the orbital period ($P_{\rm b}$), $\dot{P}_{\rm b, Gal}$ is the rate of the change of the orbital period caused by the gravitational field of the Galaxy, $\dot{P}_{\rm b, Shk} = 2.42925 \times 10^{-21} \, d_{\rm kpc} \, \mu_{T, {\rm mas \, yr^{-1}}}^2 \, P_{\rm b} ~ ~{\rm ss^{-1}}$ is the rate of the change of the orbital period caused by the proper motion of the pulsar (Shklovskii effect, \citet{shk70}), and $\dot{P}_{\rm b, int}$ is the intrinsic rate of change of the orbital period. $\mu_{T}$ is the total proper motion in milliarcsecond per year and $d_{\rm kpc}$ is the distance of the pulsar from the solar system barycentre. Similarly, one can write $\dot{P}_{\rm s, int} = \dot{P}_{\rm s, obs} - \dot{P}_{\rm s, Gal} - \dot{P}_{\rm s, Shk}$ where $\dot{P}_{\rm s, obs}$ is the observed rate of the change of the spin period, $\dot{P}_{\rm s, Gal}$ is the rate of the change of the spin period caused by the gravitational field of the Galaxy, $\dot{P}_{\rm s, Shk} = 2.42925 \times 10^{-21} \, d_{\rm kpc} \, \mu_{T, {\rm mas \, yr^{-1}}}^2 \, P_{\rm s} ~ ~{\rm ss^{-1}}$ is the rate of the change of the spin period caused by the proper motion of the pulsar (Shklovskii effect, \citet{shk70}), and $\dot{P}_{\rm s, int}$ is the intrinsic rate of change of the spin period. Here, we have ignored additional effects like the cluster potential, attraction by near-by stars, etc.

We first focus on the pulsars for which the observed period derivative are of the opposite signs than those expected and see whether we get the correct signs after eliminating the dynamical effects. We also mention the cases where the elimination of the dynamical effects reveals an intrinsic period derivative value with sign opposite to that expected. Moreover, we try to seek reasons behind those atypical values.

We exclude the pulsars in Globular clusters as their accelerations are affected by the local gravitational effects, e.g., the average cluster potential, the gravitational force from one or more near-by star(s), etc. We also exclude the pulsars belonging to either the Large Magellanic Cloud or the Small Magellanic Cloud. Since these are Milky Way's satellite galaxies, the model for the Milky Way's potential will not be valid there and one would require to model the local potential to understand the dynamics of the pulsars there. 
 
We also probe different methods of distance estimation. The most accurate distance measurement method is the use of the parallax measurements. If the parallax value for a pulsar is available, say $\Pi$ milliarcseconds (mas), then the distance to that pulsar is given by the inverse of that parallax value, in kiloparsecs (kpc). The distance values obtained in such a manner is denoted by $d_{\Pi,\rm LK} (= 1/\Pi)$. We also account for the systematic error in parallax measurements, i.e., the Lutz-Kelker (LK) bias \citep{lk73} following the algorithm provided by \citet{vdm10} available at {\tt http://psrpop.phys.wvu.edu/LKbias/index.php}. However, not all pulsars have a measured parallax value available. For such cases, one of the methods to perform distance estimation is by using the dispersion measure (DM) value and a model for electron density of cold ionized plasma, i.e., the interstellar medium. For this DM based distance estimation, we use both of the popular models for the electron density, i.e., the NE2001 model \citep{cl02,cl03}, and the YMW16 model \citep{ymw17}. Whenever we have the measurements of proper motions and some estimation of $\dot{P}_{\rm b, int}$ available, we can also use Eq. (\ref{eq:pbint1}) to estimate distance from $\dot{P}_{\rm b, Shk}$ value and compare with other methods of distance estimation aforementioned.

We use `GalDynPsr', a python based package\footnote{\tt https://github.com/pathakdhruv/GalDynPsr}, to estimate $\dot{P}_{\rm b, Gal}$, $\dot{P}_{\rm b, Shk}$, $\dot{P}_{\rm s, Gal}$, $\dot{P}_{\rm s, Shk}$, $\dot{P}_{\rm b, int}$, and $\dot{P}_{\rm s, int}$. Specifically, we use the Model-La prescribed in GalDynPsr as it is based on a fairly realistic picture of the Milky Way gravitational potential \citep{bovy15}. This model does not provide uncertainties. We, however, estimate uncertainties by employing Monte-Carlo simulation. GalDynPsr requires the Galactic longitude ($l$), Galactic latitude ($b$), distance (d), proper motion values, period and period derivative. Timing solutions are usually obtained involving either the equatorial coordinates (right ascension and declination) or the ecliptic coordinates (ecliptic longitude and ecliptic latitude). We use these values (with reported uncertainties) and use the {\tt SkyCoord} module of the astropy package to obtain the mean and the uncertainties in $l$ and $b$. For the parallax based distances, we take the uncertainties mentioned in the respective references. Among the cases where the distance estimation is done based on the DM and the electron density model of the plasma, NE2001 model calculates the uncertainties in distance values based on 20\% uncertainty in the DM. We use the same technique to calculate uncertainties in distance values in the YMW16 model. Rest of the parameters and their uncertainties are used from their respective references. We then simulate 50000 instances of the input parameters to obtain the mean and the standard deviation of $\dot{P}_{\rm b, Gal}$ (or $\dot{P}_{\rm s, Gal}$, as per the case). In the text, we mention the uncertainties, in addition to the mean values, for the values of the subsequently calculated various period derivative terms- $\dot{P}_{\rm b, Gal}$, $\dot{P}_{\rm b, Shk}$, $\dot{P}_{\rm s, Gal}$, $\dot{P}_{\rm s, Shk}$, $\dot{P}_{\rm b, int}$, and $\dot{P}_{\rm s, int}$ (as the case may be). However, for simplicity, we only mention the mean values in the tables and not the uncertainties.

\section{Correcting $\dot{P}_{\rm b, obs}$ and looking for problematic $\dot{P}_{\rm b, int}$}

First we investigate potential interesting cases where $\dot{P}_{\rm b, obs}$ might be significantly different than $\dot{P}_{\rm b, int}$. As already mentioned, $\dot{P}_{\rm b, int}$ is usually expected to be negative as it is the manifestation of the loss of the orbital energy by radiation of the gravitational waves. In such cases, $\dot{P}_{\rm b, int}$ is equivalent to $\dot{P}_{\rm b, GW}^Q$ where $\dot{P}_{\rm b, GW}^Q$ is the rate of change of the orbital period due to the emission of the quadrupolar gravitational waves, which is the only type of gravitational waves allowed under general relativity.
 
For some pulsars, however, the best fit timing solutions lead to positive values of $\dot{P}_{\rm b, obs}$. 
The version $1.60$ of the ATNF\footnote{http://www.atnf.csiro.au/research/pulsar/psrcat/} pulsar catalogue \citep{mhth05} reports 17 such pulsars in the Galactic field, three more are reported in the Version-2 data release of the International Pulsar Timing Array (IPTA) \citep{vlh16}, and one more is reported in \citet{zsd15}\footnote{This pulsar, PSR J1713$+$0747, although is a part of the IPTA program, none of the PTA solutions report $\dot{P}_{\rm b, obs}$.}. Additionally, 10 of the first 17 pulsars are also being monitored by various Pulsar Timing Array (PTA) experiments, e.g., The North American Gravitational Wave Observatory (NANOGrav), European Pulsar Timing Array (EPTA), Parkes Pulsar Timing Array (PPTA) and as a result the combined IPTA. Whenever we find a PTA timing solution, we use that, except in the case of PSR J2234+0611, where we use the timing solution given in \citet{st19} as it is based on a longer time span of observations than the public domain data of NANOGrav. Again, for the pulsars being monitored by more than one PTA, we use the timing solution obtained using the data that covers a longer time-baseline. From the above mentioned 21 pulsars, we exclude the redback PSR J1048+2339 \citep{drc16} and the black-widows PSR J0023$+$0923 \citep{abb18} and PSR J1959$+$2048 \citep{aft94}, as for all of them, higher derivatives of the orbital period that are strongly covariant with the proper motion, have been fitted in the timing solutions. There is another black-widow, PSR J0636$+$5128, for which higher derivatives of the orbital period are not fitted, but still we exclude this system as tidal effects of the companion might be responsible for the orbital variability \citep{abb18}. For the same reason, we also exclude PSR J1957$+$2516, as \citet{st16} suggests it to be either a black-widow or a redback system, as well as, PSR J2115$+$5448, another black-widow with no proper motion and parallax measurements \citep{sa16}. 

We investigate the remaining 15 pulsars with positive $\dot{P}_{\rm b, obs}$ in details in the next subsection. We divide these 15 pulsars in three categories: (i) positive $\dot{P}_{\rm b, obs}$ with both parallax and proper motion measured, (ii) positive $\dot{P}_{\rm b, obs}$ with proper motion measured but no parallax measurement, and (iii) positive $\dot{P}_{\rm b, obs}$ with measured parallax but no proper motion measurement. We did not find any pulsars in the third category. We discuss category (i) cases in subsection 2.1, and category (ii) cases in subsection 2.2. For some pulsars, we have also used the distance estimates obtained from the recent second data release of \textit{Gaia} mission (\textit{Gaia} DR2). For the benefit of the readers, such pulsar names are marked with $\ast$s.

We have also checked all the source, i.e. the ATNF catalogue and the PTA data releases whether there is any pulsar with a measured proper motion for which $\dot{P}_{\rm b, obs}$ is negative but $\dot{P}_{\rm b, int}$ becomes positive, but we did not find any such case.

\subsection{Positive $\dot{P}_{\rm b, obs}$ with both parallax and proper motion measured}

There are 13 pulsars with positive $\dot{P}_{\rm b, obs}$ and with measurements of parallax and proper motions. Six of these also have measurements of post-Keplerian (PK) orbital parameters so that we can obtain the values of the masses for the pulsar and the companion and hence calculate $\dot{P}_{\rm b, GW}^{Q}$ \citep{PCP64}. These pulsars and various parameters are listed in Table \ref{tb:positivePbdot1a}.

\begin{table*}
\caption{Different parameters, dynamical terms and $\dot{P}_{\rm b, int}$ values for the pulsars with positive $\dot{P}_{\rm b, obs}$, known parallax, known proper motion, and known sufficient PK orbital parameters to calculate $\dot{P}_{\rm b, GW}^{Q}$. PSR J1713$+$0747 is NANOGrav, EPTA as well as IPTA pulsar, however we have used timing solutions provided by \citet{zsd15} as it had the longest time span of observations. We have performed LK bias corrections to the parallax for PSR J1614$-$2230, PSR J1713$+$0747, PSR J1909$-$3744, PSR J2222$-$0137, and PSR J2234$+$0611. \citet{vlh16} already provided the LK bias corrected distance for J0437$-$4715 corresponding to the IPTA data release. For LK corrections refer to {\tt http://psrpop.phys.wvu.edu/LKbias/index.php} and \citet{vdm10}. The rows indicate the orbital period ($P_{\rm b}$), the eccentricity (Ecc.), the total proper motion ($\mu_T$), the LK-bias corrected distance obtained from parallax ($d_{\Pi,\rm LK}$), the observed rate of change of the orbital period ($\dot{P}_{\rm b, obs}$), the rate of change of the orbital period caused by the gravitational field of the Galaxy ($\dot{P}_{\rm b, Gal}$), the proper motion contribution to the rate of change of the orbital period based on $d_{\Pi,\rm LK}$ ($\dot{P}_{\rm b, Shk}$), the calculated intrinsic rate of change of the orbital period derivative ($\dot{P}_{\rm b, int}$) using Eqn (\ref{eq:pbint1}), the proper motion contribution to the rate of change of the orbital period ($\dot{P}_{\rm b, Shk, new}$) obtained from $\dot{P}_{\rm b, GW}^{Q}$ using Eqn (\ref{eq:findShk}), the distance based on $\dot{P}_{\rm b, Shk, new}$ ($d_{\rm Shk}$), the difference between $\dot{P}_{\rm b, int}$ and $\dot{P}_{\rm b, GW}^{Q}$ ($\dot{P}_{\rm b, extra}$), and the references. We do not report the uncertainties here, only the mean values of the simulations are reported.}
\begin{tabular}{l@{\hskip2pt} @{\hskip2pt}l@{\hskip2pt} @{\hskip2pt}l@{\hskip2pt} @{\hskip2pt}l@{\hskip2pt} @{\hskip2pt}l@{\hskip2pt} @{\hskip2pt}l@{\hskip2pt} @{\hskip2pt}l}
\hline \hline
Parameters & J0437$-$4715 & J1614$-$2230 & J1713$+$0747 & J1909$-$3744 & J2222$-$0137 & J2234$+$0611\\ 
\hline
$P_{\rm b}$ & 5.7410 & 8.6866 & 67.8251 & 1.5334 & 2.4458 & 32.0014\\
 (days)   & & & & & & \\ \\
Ecc. & $1.9181\times10^{-5}$ & $1.3362\times10^{-6}$ & $7.4940\times10^{-5}$ & $1.1580\times10^{-7}$ & $3.8097\times10^{-4}$ & $0.1293$\\ \\
$\mu_T$  & 140.914 & 32.7 & 6.2830 & 37.01 & 45.09 & 27.10 \\
(mas/yr) & & & & & & \\ \\
$d_{\Pi,\rm LK}$ & 0.156 & 0.662 & 1.144 & 1.086 & 0.267& 0.966 \\
(kpc) & & & & & & \\ \\
$\dot{P}_{\rm b, obs}$ & $3.724\times 10^{-12}$ & $1.69\times 10^{-12}$ & $3.6\times 10^{-13}$ & $5.022\times10^{-13}$ & $2.0\times10^{-13}$ & $1.8\times 10^{-12}$\\
(${\rm s \, s^{-1}}$) & & & & & & \\ \\
$\dot{P}_{\rm b, Gal}$ & $-2.23\times 10^{-14}$ & $6.17\times10^{-15}$ & $-3.21\times 10^{-13}$ & $3.61\times10^{-15}$ & $-1.67\times10^{-14}$ & $-4.71\times 10^{-13}$\\
 (${\rm s \, s^{-1}}$) & & & & & & \\ \\
 $\dot{P}_{\rm b, Shk}$ & $3.73\times 10^{-12}$ & $1.29\times 10^{-12}$ & $6.43\times 10^{-13}$ & $4.79\times10^{-13}$ &  $2.79\times10^{-13}$ & $4.77\times 10^{-12}$\\
 (${\rm s \, s^{-1}}$)   & & & & & & \\ \\
 $\dot{P}_{\rm b, int}$  & $1.35\times 10^{-14}$ & $3.92\times 10^{-13}$ & $3.9\times10^{-14}$ & $1.95\times10^{-14}$ & $-6.2\times10^{-14}$ & $-2.5\times 10^{-12}$\\
 (${\rm s \, s^{-1}}$)   & & & & & & \\ \\
 $\dot{P}_{\rm b, Shk, new}$ & $3.75\times10^{-12}$ & $1.68\times 10^{-12}$ & $6.8\times10^{-13}$ & $5.01\times10^{-13}$ & $2.2\times10^{-13}$ & $2.3\times10^{-12}$\\
 (${\rm s \, s^{-1}}$)   & & & & & & \\ \\
 $d_{\rm Shk}$ & 0.16 & 0.86 & 1.21 & 1.14 & 0.22 & 0.46\\ 
(kpc) & & & & & & \\ \\ 
$\dot{P}_{\rm b, extra}$ & $1.38\times10^{-14}$ & $3.93\times10^{-13}$ & $3.9\times10^{-14}$ & $2.22\times10^{-14}$ & $-5.4\times10^{-14}$ & $-2.5\times10^{-12}$\\
 (${\rm s \, s^{-1}}$)   & & & & & & \\ \\
Reference & IPTA & NANOGrav & \citet{zsd15} & NANOGrav & \citet{cfg17}& \citet{st19} \\
\hline \hline
\label{tb:positivePbdot1a}
\end{tabular}
\end{table*}

For the sake of consistency check, we use the following expression that assumes $\dot{P}_{\rm b, int}$ to be equivalent to $\dot{P}_{\rm b, GW}^{Q}$:
\begin{equation}
\dot{P}_{\rm b, Shk, new} = \dot{P}_{\rm b, obs} - \dot{P}_{\rm b, Gal} -\dot{P}_{\rm b, GW}^{Q}  
\label{eq:findShk}
\end{equation} This $\dot{P}_{\rm b, Shk, new}$ yields a value of the distance ($d_{\rm Shk}$). This distance can be compared with the estimates of distances from other methods. Ideally, the distance estimates should match for various methods. This is possible only when enough post-Keplerian parameters have been measured to estimate the values of the pulsar and the companion masses that can be used to calculate $ \dot{P}_{\rm b, GW}^{Q} $. Additionally, if the measured value of $\dot{P}_{\rm b, obs}$ is only an upper limit or the existence of additional dynamical effects are not ruled out with confidence, Eq. (\ref{eq:findShk}) gives the upper limit of $\dot{P}_{\rm b, Shk}$ and hence, the upper limit of the distance. Whenever possible, we also calculate $\dot{P}_{\rm b, extra} = \dot{P}_{\rm b, int} - \dot{P}_{\rm b, GW}^{Q}$ where $\dot{P}_{\rm b, int}$ is estimated using Eqn. (\ref{eq:pbint1}) and $ \dot{P}_{\rm b, GW}^{Q}$ from the expression given by \citet{PCP64}.

 Now we discuss these anomalous pulsars in detail. The results are also summarised in Tables \ref{tb:positivePbdot1a} and \ref{tb:positivePbdot1b}.

\vskip 0.1cm

\textbf{PSR J0437$-$4715}$^*$: 

PSR J0437$-$4715 has $\dot{P}_{\rm b, obs}= 3.724(5)\times10^{-12}~{\rm s s^{-1}}$. Using $d_{\Pi,\rm LK}=0.156(1)$ kpc, we get $\dot{P}_{\rm b, Gal} = -2.23(1)\times10^{-14}~{\rm s s^{-1}}$ and $ \dot{P}_{\rm b, Shk} = 3.73(2)\times 10^{-12}~{\rm s s^{-1}}$ giving $\dot{P}_{\rm b, int} = 1.4(24)\times 10^{-14}~{\rm s s^{-1}}$ (Eqn. ((\ref{eq:pbint1})). It is an interesting fact that the value of $\dot{P}_{\rm b, Shk}$ is very close to the value of $\dot{P}_{\rm b, obs}$ making the value of $\dot{P}_{\rm b, int}$ very sensitive on the value of the distance used.

The IPTA timing solution provides the measured value of two PK parameters, which are, $\dot{\omega}$, the rate periastron advance and $\sin i$, where $i$ is the angle of inclination of the orbit. Using the analytical expressions for these terms \citep{lk05} along with the values of the eccentricity, the projected semi-major axis, and the orbital period, we get the masses of the pulsar and the companion as $1.21\,{\rm M_{\odot}}$ and $0.19 \, {\rm M_{\odot}}$, respectively. Using these masses, we get $\dot{P}_{\rm b, GW}^{Q} = -2.48\times10^{-16}~{\rm s s^{-1}}$. This leads to $\dot{P}_{\rm b, extra} = 1.38\times10^{-14}~{\rm s s^{-1}}$

Using these values of $\dot{P}_{\rm b, Gal}$ and $\dot{P}_{\rm b, GW}^{Q}$ in Eqn. (\ref{eq:findShk}), we get $ \dot{P}_{\rm b, Shk, new} = 3.75\times10^{-12}~{\rm s s^{-1}}$ and $d_{\rm Shk} = 0.16$ kpc.

If we use the recent distance measurement of this pulsar ($156.77$ pc) based on the second data release of \textit{Gaia} mission (\textit{Gaia} DR2) \citep{ming18}, and other parameters from the IPTA timing solution, we get $\dot{P}_{\rm b, int} = -4.79\times10^{-15}~{\rm s s^{-1}}$.

Earlier, \citet{rhc16} used the Eq. (\ref {eq:findShk}) to obtain $ \dot{P}_{\rm b, Shk, new} = 3.7513 \times 10^{-12}~{\rm s s^{-1}}$ and $d_{\rm Shk} = 0.1568$ kpc, using $\dot{P}_{\rm b, obs} = 3.728 \times 10^{-12}~{\rm s s^{-1}}$, $\mu_T = 140.9115$ mas/yr, $\dot{P}_{\rm b, Gal} = -2.3 \times 10^{-14}~{\rm s s^{-1}}$, and $\dot{P}_{\rm b, GW}^{Q} = -3.2\times10^{-16}~{\rm s s^{-1}}$.

\vskip 0.1cm

\textbf{PSR J1614$-$2230}: Using $d_{\Pi,\rm LK} = 0.66(5)$ kpc, we obtain $\dot{P}_{\rm b, Gal} = 6.2(38)\times10^{-15}~{\rm s s^{-1}}$ and $ \dot{P}_{\rm b, Shk} = 1.29(11)\times 10^{-12}~{\rm s s^{-1}}$ giving $ \dot{P}_{\rm b, int} = 3.9(24)\times 10^{-13}~{\rm s s^{-1}}$ (Eqn. \ref{eq:pbint1}). 

We obtained the mass of the pulsar ($1.91\,{\rm M_{\odot}}$) using the values of PK parameters, $\sin i$ ($0.9999$) and the mass of the companion ($0.49\,{\rm M_{\odot}}$), provided in the NANOGrav timing solution. Using these masses we calculate the value of $\dot{P}_{\rm b, GW}^{Q} = -4.16\times10^{-16}~{\rm s s^{-1}}$. Interestingly, this value of $\dot{P}_{\rm b, GW}^{Q}$ differs significantly from the $\dot{P}_{\rm b, int}$ resulting in the value of $\dot{P}_{\rm b, extra} = 3.93\times10^{-13}~{\rm s s^{-1}}$.

Using these values of $\dot{P}_{\rm b, Gal}$ and $\dot{P}_{\rm b, GW}^{Q}$ in Eqn. (\ref{eq:findShk}), we get $ \dot{P}_{\rm b, Shk, new} = 1.68\times10^{-12}~{\rm s s^{-1}}$ and $d_{\rm Shk} = 0.86$ kpc. This $d_{\rm Shk}$ differs by $30.4\%$ from $d_{\Pi,\rm LK}$. This implies that if $d_{\Pi,\rm LK}$ is the correct distance, then there must be additional factors affecting $\dot{P}_{\rm b, Gal}$.

\vskip 0.1cm

\textbf{PSR J1713$+$0747}: We use $d_{\Pi,\rm LK}= 1.14(4)$ kpc to obtain $\dot{P}_{\rm b, Gal} = -3.21(12)\times10^{-13}~{\rm s s^{-1}}$, $\dot{P}_{\rm b, Shk} = 6.43(21)\times10^{-13}~{\rm s s^{-1}}$, and consequently $\dot{P}_{\rm b, int} = 3.9(172)\times10^{-14}~{\rm s s^{-1}}$.

\citet{zsd15} gives values of two PK parameters, $\sin i$ $(0.9505)$ and $\dot{\omega}$ (0.00024 deg/yr). Using these values, we calculate the values of the mass of the pulsar ($1.38\, {\rm M_{\odot}}$) and the mass of the companion ($0.29\, {\rm M_{\odot}}$). Using the values of the masses, we get $\dot{P}_{\rm b, GW}^{Q} = -6.41\times10^{-18}~{\rm s s^{-1}}$. Here too, we see that $\dot{P}_{\rm b, GW}^{Q}$ differs significantly from $\dot{P}_{\rm b, int}$ with $\dot{P}_{\rm b, extra} = 3.86\times10^{-14}~{\rm s s^{-1}}$.

Using the values of $\dot{P}_{\rm b, Gal}$ and $\dot{P}_{\rm b, GW}^{Q}$ in Eqn. (\ref{eq:findShk}), we get $\dot{P}_{\rm b, Shk, new} = 6.81\times10^{-13}~{\rm s s^{-1}}$ and $d_{\rm Shk} = 1.21$ kpc which differs by $6.0\%$ from $d_{\Pi,\rm LK}$.

Earlier, \citet{zsd15} used Eq. 5 in \citet{nt95} and Eq. 11 in \citet{lwz09} to obtain $\dot{P}_{\rm b, Gal}$ as $-0.10\times10^{-12}~{\rm s s^{-1}}$ and calculated $\dot{P}_{\rm b, Shk}$ to be $0.65\times10^{-12}~{\rm s s^{-1}}$. They obtained $\dot{P}_{\rm b, int} = -0.20\times10^{-12}~{\rm s s^{-1}}$ and $\dot{P}_{\rm b, GW}^{Q} = -6\times10^{-18}~{\rm s s^{-1}}$. They, however, did not correct the parallax measurement and the distance for LK bias. 

\vskip 0.1cm

\textbf{PSR J1909$-$3744}: Using $d_{\Pi,\rm LK} = 1.09(5)$ kpc, we obtain $\dot{P}_{\rm b, Gal} = 3.62(66)\times10^{-15}~{\rm s s^{-1}}$ and $ \dot{P}_{\rm b, Shk} = 4.79(22)\times 10^{-13}~{\rm s s^{-1}}$ giving $\dot{P}_{\rm b, int} = 1.9(22)\times 10^{-14}~{\rm s s^{-1}}$ (Eqn. \ref{eq:pbint1}).

We obtained the mass of the pulsar ($1.48\,{\rm M_{\odot}}$) using the values of PK parameters, $\sin i$ ($0.99808$) and the mass of the companion ($0.2077\,{\rm M_{\odot}}$), provided in the NANOGrav timing solution. Using these masses we calculate the value of $\dot{P}_{\rm b, GW}^{Q} = -2.76\times10^{-15}~{\rm s s^{-1}}$. The value of $\dot{P}_{\rm b, GW}^{Q}$ here too, differs significantly from that of $\dot{P}_{\rm b, int}$ and this leads to $\dot{P}_{\rm b, extra} = 2.22\times10^{-14}~{\rm s s^{-1}}$.

Using these values of $\dot{P}_{\rm b, Gal}$ and $\dot{P}_{\rm b, GW}^{Q}$ in Eqn. (\ref{eq:findShk}), we get $ \dot{P}_{\rm b, Shk, new} = 5.01\times10^{-13}~{\rm s s^{-1}}$ and $d_{\rm Shk} = 1.14$ kpc. This $d_{\rm Shk}$ differs by $4.6\%$ from $d_{\Pi,\rm LK}$. 

Earlier, \citet{rhc16} calculated $\dot{P}_{\rm b, GW}^{Q} = -2.7\times10^{-15}~{\rm s s^{-1}}$ and the mass of the pulsar as $1.47\, {\rm M_{\odot}}$ using the measured value of mass of the companion, $0.2067\, {\rm M_{\odot}}$. They used these values and the proper motion of the pulsar to get a distance of $1.140$ kpc.

\vskip 0.1cm

\textbf{PSR J2222$-$0137}: Using $d_{\Pi,\rm LK} = 0.27(3)$ kpc, we obtain $\dot{P}_{\rm b, Gal} = -1.67(13)\times10^{-14}~{\rm s s^{-1}}$ and $ \dot{P}_{\rm b, Shk} = 2.79(31)\times 10^{-13}~{\rm s s^{-1}}$ giving $ \dot{P}_{\rm b, int} = -6.2(95)\times 10^{-14}~{\rm s s^{-1}}$ (Eqn. \ref{eq:pbint1}). For PSR J2222$-$0137, hence, we get a negative value of $\dot{P}_{\rm b, int}$, as expected.

We use the value of $\dot{P}_{\rm b, GW}^{Q}$ $(-7.7\times10^{-15}~{\rm s s^{-1}})$, as reported in \citet{cfg17}, and obtain $\dot{P}_{\rm b, extra} = -5.4\times10^{-14}~{\rm s s^{-1}}$.

Using these values of $\dot{P}_{\rm b, Gal}$ and $\dot{P}_{\rm b, GW}^{Q}$ in Eqn. (\ref{eq:findShk}), we get $ \dot{P}_{\rm b, Shk, new} = 2.24\times10^{-13}~{\rm s s^{-1}}$ and $d_{\rm Shk} = 0.22$ kpc which is less by $19.5\%$ from $d_{\Pi,\rm LK}$.

Earlier, \citet{cfg17} estimated $\dot{P}_{\rm b, int} = -0.063\times10^{-12} ~{\rm s s^{-1}}$ which is close to our GalDynPsr result. They also calculated, as mentioned above, $\dot{P}_{\rm b, GW}^{Q} = -7.7\times10^{-15}~{\rm s s^{-1}}$. On subtracting the expected GR contribution from $\dot{P}_{\rm b, int}$, they obtained an excess $\dot{P}_{\rm b, extra} = -5.5\times10^{-14}~{\rm s s^{-1}}$, which they claimed, represents the upper limit for dipolar GW emission.

\vskip 0.1cm

\textbf{PSR J2234+0611}: For the orbital parameters, we use the timing solutions derived using the DDGR model as given in Table 2 of \citet{st19}\footnote{\citet{st19} gave $\dot{P}_{\rm b, GW}^{Q} = 2.62\times10^{-17}~{\rm s s^{-1}}$ where the minus sign is missing.}. We specifically choose to use DDGR model's timing solutions as it provides the maximum number of PK parameters among the other models in the Table 2 of \citet{st19}. From this data we use $\dot{P}_{\rm b, obs} = 1.8(25)\times10^{-12} ~{\rm s s^{-1}}$. Using $d_{\Pi,\rm LK} = 0.97(5)$ kpc, we obtain $\dot{P}_{\rm b, Gal} = -4.71(9)\times10^{-13}~{\rm s s^{-1}}$ and $ \dot{P}_{\rm b, Shk} = 4.77(25)\times 10^{-12}~{\rm s s^{-1}}$ giving $\dot{P}_{\rm b, int} = -2.5(25)\times 10^{-12}~{\rm s s^{-1}}$ (Eqn. \ref{eq:pbint1}). Here too we get a negative values of $\dot{P}_{\rm b, int}$, as expected.

Using the values of periastron advance, $\dot{\omega} = 8.863\times10^{-4}~{\rm deg\, yr^{-1}}$, and mass of companion, $ M_{C} = 0.003\, {\rm M_{\odot}}$, we obtain mass of the pulsar, $M_{P} = 1.38\, {\rm M_{\odot}}$ and subsequently, $\dot{P}_{\rm b, GW}^{Q} = -2.62\times10^{-17}~{\rm s s^{-1}}$. This value of $\dot{P}_{\rm b, GW}^{Q}$ is five orders of magnitude smaller than $\dot{P}_{\rm b, int}$ obtained using Eqn.\ref{eq:pbint1}. Using this value, we further obtain, $\dot{P}_{\rm b, extra} = -2.5\times10^{-12}~{\rm s s^{-1}}$. 

Using these values of $\dot{P}_{\rm b, Gal}$, and $\dot{P}_{\rm b, GW}^{Q}$ in Eqn. (\ref{eq:findShk}), we get $ \dot{P}_{\rm b, Shk, new} = 2.27\times10^{-12}~{\rm s s^{-1}}$ and $d_{\rm Shk} = 0.46$ kpc which is $52.36\%$ smaller than $d_{\Pi,\rm LK}$.

\vskip 0.2cm

There are seven more pulsars with positive $\dot{P}_{\rm b, obs}$ and with measurements of parallax and proper motions. However, for these pulsars, measurements of enough PK parameters are unavailable till now. These pulsars are reported in Table \ref{tb:positivePbdot1b}. For two of these pulsars, we obtain negative values of $\dot{P}_{\rm b, int}$. These pulsars are PSR J1012$+$5307 and PSR J1022$-$1001. For PSR J1012$+$5307, we used the EPTA timing solution provided by \citet{dcl16}. For PSR J1022$-$1001, we used the LK bias corrected distance from \citet{vlh16} and other parameters from the IPTA par file of the pulsar. For PSR J1302$-$6350, since, apart from using distance based on parallax, we also calculate dynamical terms using distance based on optical measurements, we display the results in a separate table (Table \ref{tb:positivePbdot1bJ1302}).

\begin{table*}
\caption{Different parameters, dynamical terms and $\dot{P}_{\rm b, int}$ values for the pulsars with positive $\dot{P}_{\rm b, obs}$, known parallax and known proper motion. We have used Lutz-Kelker (LK) bias corrections to the parallax for PSR J1603$-$7202. \citet{dcl16} already provides the LK bias corrected distances for PSR J0613$-$0200 and PSR J1012$+$5307. Similarly, \citet{vlh16} provides the LK bias corrected distances for the IPTA pulsars PSR J0900$-$3144, PSR J1022$-$1001, and PSR J2145$-$0750. For LK corrections refer to {\tt http://psrpop.phys.wvu.edu/LKbias/index.php} and \citet{vdm10}. Due to the lack of sufficient PK parameters, $\dot{P}_{\rm b, GW}^{Q}$ values cannot be calculated for these pulsars. See caption of Table \ref{tb:positivePbdot1a} for description of rows. We do not report the uncertainties here, only the mean values of the simulations are reported.}
\begin{tabular}{l@{\hskip2pt} @{\hskip2pt}l@{\hskip2pt} @{\hskip2pt}l@{\hskip2pt} @{\hskip2pt}l@{\hskip2pt} @{\hskip2pt}l@{\hskip2pt} @{\hskip2pt}l@{\hskip2pt} @{\hskip2pt}l}
\hline \hline
Parameters & J0613$-$0200 & J0900$-$3144 & J1012$+$5307 & J1022$+$1001 & J1603$-$7202 & J2145$-$0750 \\ 
\hline
$P_{\rm b}$ & 1.1985 & 18.7376 & 0.6047 & 7.8051 & 6.3086 & 6.8389\\
 (days)    & & & & & & \\ \\
%Ecc. & & & & & & \\ \\
$\mu_T$  & 10.514 & 2.0658 & 25.615 & 19.7186 & 7.73 & 13.0765\\
(mas/yr) & & & & & &  \\ \\
$d_{\Pi,\rm LK}$ & 0.777 & 0.5474 & 1.148 & 0.52 & 0.5405 & 0.57\\ % 0.54741 & 1.148 & 0.52 & 0.54048
(kpc) & & & & & & \\ \\
$\dot{P}_{\rm b, obs}$ &  $4.8\times 10^{-14}$  & $1.18\times 10^{-11}$ & $6.1\times10^{-14}$ & $2.19\times 10^{-13}$ & $3.10\times10^{-13}$ & $8.66\times 10^{-14}$\\
(${\rm s \, s^{-1}}$) & & & & & & \\ \\
$\dot{P}_{\rm b, Gal}$ & $3.18\times10^{-15}$ & $-7.06\times 10^{-14}$  & $-3.90\times10^{-15}$ & $-7.18\times 10^{-14}$ & $-9.84\times10^{-15}$ & $-6.69\times 10^{-14}$ \\
 (${\rm s \, s^{-1}}$) & & & & & & \\ \\
 $\dot{P}_{\rm b, Shk}$ & $2.17\times 10^{-14}$ & $9.27\times 10^{-15}$ & $9.58\times10^{-14}$ & $3.31\times 10^{-13}$ & $4.22\times10^{-14}$ & $1.40\times 10^{-13}$\\
 (${\rm s \, s^{-1}}$)   & & & & & & \\ \\
 $\dot{P}_{\rm b, int}$  & $2.3\times 10^{-14}$  & $1.19\times 10^{-11}$ & $-3.1\times10^{-14}$ & $-4.01\times 10^{-14}$ & $2.78\times10^{-13}$ & $1.35\times 10^{-14}$ \\
 (${\rm s \, s^{-1}}$)   & & & & & & \\ \\
 $\dot{P}_{\rm b, GW}^{Q} + \dot{P}_{\rm b, Shk}$ & $4.5\times10^{-14}$ & $1.19\times10^{-11}$ & $6.5\times10^{-14}$ & $2.91\times10^{-13}$ & $3.20\times10^{-13}$ & $1.53\times10^{-13}$ \\
 (${\rm s \, s^{-1}}$)   & & & & & & \\ \\
Reference & \footnotesize{\citet{dcl16}} & ~\footnotesize{IPTA} & \footnotesize{\citet{dcl16}} & ~\footnotesize{IPTA} & \footnotesize{\citet{rhc16}} & ~\footnotesize{IPTA}\\
\hline \hline
\label{tb:positivePbdot1b}
\end{tabular}
\end{table*}

\begin{table*}
\caption{Different parameters, dynamical terms and $\dot{P}_{\rm b, int}$ values for PSR J1302$-$6350, possessing positive $\dot{P}_{\rm b, obs}$, known parallax and known proper motion. Due to the lack of sufficient PK parameters, $\dot{P}_{\rm b, GW}^{Q}$ values cannot be calculated for this pulsar too. The columns indicate the pulsar name, the orbital period ($P_{\rm b}$), the total proper motion ($\mu_T$), the observed rate of change of the orbital period ($\dot{P}_{\rm b, obs}$), the rate of change of the orbital period caused by the gravitational field of the Galaxy ($\dot{P}_{\rm b, Gal}$), proper motion contribution to the rate of change of orbital period ($\dot{P}_{\rm b, Shk}$), calculated intrinsic rate of change of orbital period derivative ($\dot{P}_{\rm b, int}$), and the references. We do not report the uncertainties here, only the mean values of the simulations are reported. Note that, as the magnitude of the value of $\dot{P}_{\rm b, obs}$ is much larger than the magnitude of both $\dot{P}_{\rm b, Gal}$ and $\dot{P}_{\rm b, Shk}$, with the present accuracy of the parameters, we do not see any difference in the value of $\dot{P}_{\rm b, GW}^{Q} + \dot{P}_{\rm b, Shk}$ from that of $\dot{P}_{\rm b, obs}$.}
\begin{tabular}{l@{\hskip1pt} @{\hskip1pt}l l l@{\hskip1pt} @{\hskip1pt}l@{\hskip2pt} @{\hskip2pt}l@{\hskip2pt} @{\hskip1pt}l@{\hskip1pt} @{\hskip1pt}l}
\hline \hline
 Pulsar & $P_{\rm b}$ & $\mu_T$ & $\dot{P}_{\rm b, obs}$ & $\dot{P}_{\rm b, Gal}$ & $\dot{P}_{\rm b, Shk}$ &  $\dot{P}_{\rm b, int}$ & References  \\ 
 & (days) & (mas/yr) & (${\rm s \, s^{-1}}$) & (${\rm s \, s^{-1}}$) & (${\rm s \, s^{-1}}$) &  (${\rm s \, s^{-1}}$) &  \\ \hline
J1302$-$6350 & 1236.7245 & 7.93 & $1.4\times10^{-8}$ & &  &   & \footnotesize{\citet{sjm14}} \\  \\
 
for $d= 2.3$ kpc &  & &  & $-1.36\times10^{-11}$ & $3.76\times10^{-11}$ & $1.4\times10^{-8}$ &  \\
(\citet{nrh11}) &  &  & &  &  &  & \\ \\

for $d= 2.6$ kpc &  & &  & $-1.66\times10^{-11}$ & $3.34\times10^{-11}$ & $1.4\times10^{-8}$  &  \\ 
(\citet{mj18}) &  &  &  & &  &  &  \\ \\  \hline \hline
   
\label{tb:positivePbdot1bJ1302}
\end{tabular}
\end{table*}

For these seven pulsars, owing to the limited knowledge of PK parameters, we cannot calculate $\dot{P}_{\rm b, GW}^{Q}$. However, we can put a constraint on the sum of $\dot{P}_{\rm b, GW}^{Q}$ and $\dot{P}_{\rm b, Shk, new}$ by using the expression $\dot{P}_{\rm b, obs} - \dot{P}_{\rm b, Gal} =\dot{P}_{\rm b, GW}^{Q} + \dot{P}_{\rm b, Shk, new}$. If in the future, improved timing solution gives measurements of relevant PK parameters to estimate the mass of the companion and the pulsar, it will be possible to calculate the value of $\dot{P}_{\rm b, GW}^{Q}$. This value could then be subtracted from the above sum in order to get $\dot{P}_{\rm b, Shk, new}$, which could be used to get a measurement of the distance of the pulsar, which can be compared with $d_{\Pi,\rm LK}$. The results are compiled in the Tables \ref{tb:positivePbdot1b} and \ref{tb:positivePbdot1bJ1302}, and we discuss these pulsars one by one.

\vskip 0.2cm

\textbf{PSR J0613$-$0200}: PSR J0613$-$0200 is a part of the IPTA, the NANOGrav, as well as the EPTA. However, we use the EPTA timing solution given in \citet{dcl16} as it covers the longest time-baseline. Using $d_{\Pi,\rm LK}= 0.78(8)$ kpc, we obtain $\dot{P}_{\rm b, Gal} = 3.18(43)\times10^{-15}~{\rm s s^{-1}}$ and $ \dot{P}_{\rm b, Shk} = 2.17(22)\times10^{-14}~{\rm s s^{-1}}$ giving $ \dot{P}_{\rm b, int} = 2.3(11)\times10^{-14}~{\rm s s^{-1}}$ (Eqn. \ref{eq:pbint1}). Note that, after correcting for the dynamical effects, the $\dot{P}_{\rm b, int}$ value, although still positive, reduced in magnitude from the $\dot{P}_{\rm b, obs}$ value. 

\vskip 0.1cm

\textbf{PSR J0900$-$3144}: PSR J0900$-$3144 is a part of both, the IPTA and the EPTA. We use the IPTA timing solution, even though the EPTA data is based on a longer time span of observation, because only the IPTA timing solution provides a $\dot{P}_{\rm b, obs}$ value. Using $d_{\Pi,\rm LK}= 0.55(20)$ kpc, we obtain $\dot{P}_{\rm b, Gal} = -7.1(22)\times10^{-14}~{\rm s s^{-1}}$ and $ \dot{P}_{\rm b, Shk} = 9.3(38)\times10^{-15}~{\rm s s^{-1}}$ giving $ \dot{P}_{\rm b, int} = 1.19(61)\times10^{-11}~{\rm s s^{-1}}$ (Eqn. \ref{eq:pbint1}). 

For this case, the mean value of $\dot{P}_{\rm b, int}$ is slightly larger than $\dot{P}_{\rm b, obs}$, contrary to the expectation. However, if we consider the uncertainties, it is still possible to have $\dot{P}_{\rm b, int} < \dot{P}_{\rm b, obs}$. Moreover, even if $\dot{P}_{\rm b, int} > \dot{P}_{\rm b, obs}$, it could be explained by the proximity of the Gum Nebula to the pulsar. \citet{pg15} mentions that PSR J0900-3144 is close to the Nebula, and so the molecular clouds present in the nebula might exert a local acceleration on the pulsar which might be the cause of discrepancy. 

\vskip 0.1cm

\textbf{PSR J1012$+$5307}$^*$: PSR J1012$+$5307, like PSR J0613$-$0200, is also a part of all the IPTA, the NANOGrav, as well as the EPTA but we use the EPTA timing solution given in \citet{dcl16} as it covers the longest time-baseline. Using $d_{\Pi,\rm LK}= 1.15(24)$ kpc, we obtain $\dot{P}_{\rm b, Gal} = -3.9(6)\times10^{-15}~{\rm s s^{-1}}$ and $ \dot{P}_{\rm b, Shk} = 9.6(20)\times10^{-14}~{\rm s s^{-1}}$ giving $ \dot{P}_{\rm b, int} = -3.1(20)\times10^{-14}~{\rm s s^{-1}}$ (Eqn. \ref{eq:pbint1}). Thus, we can see after subtracting the dynamical effects from the positive $\dot{P}_{\rm b, obs}$ value, we obtain a negative $\dot{P}_{\rm b, int}$.

Using the recent distance measurement of this pulsar ($0.907$ kpc) based on the \textit{Gaia} DR2 \citep{ming18} and the rest of the input parameters from the EPTA timing solution given in \citet{dcl16}, we get $\dot{P}_{\rm b, int} = -1.03\times10^{-14}~{\rm s s^{-1}}$.

\vskip 0.1cm

\textbf{PSR J1022$-$1001}: PSR J1022$-$1001, like PSR J0900$-$3144, is a part of both, the IPTA and the EPTA. Here too, we use the IPTA timing solution, even though the EPTA data is based on a longer time span, because only the IPTA timing solution provides a $\dot{P}_{\rm b, obs}$ value.
Using $d_{\Pi,\rm LK}= 0.52(9)$ kpc, we obtain $\dot{P}_{\rm b, Gal} = -7.2(5)\times10^{-14}~{\rm s s^{-1}}$ and $\dot{P}_{\rm b, Shk} = 3.3(36)\times10^{-13}~{\rm s s^{-1}}$ giving $\dot{P}_{\rm b, int} = -4(38)\times10^{-14}~{\rm s s^{-1}}$ (Eqn. \ref{eq:pbint1}). In this case too, as we can see, the subtraction of the dynamical affects from the positive $\dot{P}_{\rm b, obs}$ leads to the desired negative $\dot{P}_{\rm b, int}$.

Recently, \citet{del19} gave proper motion and parallax values based on VLBI observations for 57 pulsars with PSR J1022$-$1001 being one of them. We obtain the LK bias-corrected distance form that parallax value as $0.72$ kpc. Using this distance and the proper motion values from \citet{del19}, along with the rest of the input parameters from the IPTA timing solution, we obtain $\dot{P}_{\rm b, int} = -8.80\times10^{-16}~{\rm s s^{-1}}$.

\vskip 0.1cm

\textbf{PSR J1302$-$6350}$^*$: 

PSR J1302$-$6350 has $\dot{P}_{\rm b, obs} = 1.4(7)\times10^{-8} ~{\rm s s^{-1}}$ as reported by \citet{sjm14}. For this pulsar, the preferred distance measurement as per the ATNF catalogue is $d = 2.3(4)$ kpc estimated by \citet{nrh11} using the analysis of optical spectra of the companion star as well as the analysis of the interstellar atomic lines. Since, it relies on the direct optical measurements, this distance is a better estimate than the dispersion measure distance estimate based on an electron density model. Using this distance we get $\dot{P}_{\rm b, Gal} = -1.36(37)\times10^{-11}~{\rm s s^{-1}}$ and $\dot{P}_{\rm b, Shk} = 3.8(17)\times10^{-11}~{\rm s s^{-1}}$ giving $\dot{P}_{\rm b, int} = 1.4(7)\times10^{-8}~{\rm s s^{-1}}$. Apart from the distance measurement, we have used other parameters, such as, the Galactic coordinates and the proper motion values, also from the timing solution provided in \citet{sjm14} here. 

However, \citet{mj18} calculated a systematics bias-corrected, parallax-based distance to be 2.6 kpc. They argued that this distance is fully consistent with the previous distance estimation. Using this distance we get $\dot{P}_{\rm b, Gal} = -1.66(41)\times10^{-11}~{\rm s s^{-1}}$ and $\dot{P}_{\rm b, Shk} = 3.34(51)\times10^{-11}~{\rm s s^{-1}}$ giving $\dot{P}_{\rm b, int} = 1.4(7)\times10^{-8}~{\rm s s^{-1}}$. We report these values in Table \ref{tb:positivePbdot1bJ1302} too. Here, we have used the timing solutions provided by \citet{mj18} for the pulsar position and proper motion values as well. However, we use the values of the orbital period and the orbital period derivative from \citet{sjm14}.   

Recently, \citet{jen18} reported the proper motion and the LK bias-corrected distance values based on \textit{Gaia} DR2. Using these values of the proper motion and the LK bias-corrected distance ($2.26$ kpc), along with the rest of the parameters from \citet{sjm14}, we get $\dot{P}_{\rm b, int} = 1.4\times10^{-8}~{\rm s s^{-1}}$

\citet{sjm14} argued that the positive $\dot{P}_{\rm b, int}$ of this pulsar is due to the mass loss from the companion. 

\vskip 0.1cm

\textbf{PSR J1603$-$7202}: PSR J1603$-$7202, is a IPTA pulsar but we use the timing solution provided in the \citet{rhc16} as the IPTA timing solution does not contain a measured parallax and $\dot{P}_{\rm b, obs}$, unlike \citet{rhc16}. Using $d_{\Pi,\rm LK}= 0.5(3)$ kpc, we obtain $\dot{P}_{\rm b, Gal} = -9.8(60)\times10^{-15}~{\rm s s^{-1}}$ and $ \dot{P}_{\rm b, Shk} = 4.2(24)\times10^{-14}~{\rm s s^{-1}}$ giving $ \dot{P}_{\rm b, int} = 2.8(15)\times10^{-13}~{\rm s s^{-1}}$ (Eqn. \ref{eq:pbint1}). Here, after correcting for the dynamical effects, the $\dot{P}_{\rm b, int}$ value, although still positive, reduces in magnitude from the $\dot{P}_{\rm b, obs}$ value.

\vskip 0.1cm
 
\textbf{PSR J2145$-$0750}: PSR J2145$-$0750 is also a part of the IPTA, the NANOGrav, and the EPTA. We use the IPTA timing solution for this pulsar as it provides a $\dot{P}_{\rm b, obs}$ value and is based on the longest time span of observation among the rest of the PTAs. Like PSR J1022$-$1001, \citet{del19} also gave proper motion and parallax values based on the VLBI observations for PSR J2145$-$0750. We obtain the LK bias-corrected distance form that parallax value as $0.62$ kpc. 
Using $d_{\Pi,\rm LK}= 0.57(11)$ kpc, we obtain $\dot{P}_{\rm b, Gal} = -6.69(69)\times10^{-14}~{\rm s s^{-1}}$ and $ \dot{P}_{\rm b, Shk} = 1.40(27)\times10^{-13}~{\rm s s^{-1}}$ giving $ \dot{P}_{\rm b, int} = 1.4(47)\times10^{-14}~{\rm s s^{-1}}$ (Eqn. \ref{eq:pbint1}). Here too, after correcting for the dynamical effects, the $\dot{P}_{\rm b, int}$ value remains still positive but reduces in magnitude from the $\dot{P}_{\rm b, obs}$ value. 
 
\subsection{Positive $\dot{P}_{\rm b, obs}$ with proper motion measured but no parallax measurement}

For the cases where no parallax measurement was available (PSR J1518$+$4904 and PSR J2129$-$5721), we use the dispersion measure based distances using both YMW16 and NE2001 models. Table \ref{tb:positivePbdot3a} refers to the case where the $\dot{P}_{\rm b, GW}^{Q}$ value is available whereas Table \ref{tb:positivePbdot3b} refers to the case where sufficient PK parameters are not measured and hence $\dot{P}_{\rm b, GW}^{Q}$ cannot be obtained. We discuss these in detail below.

\begin{table*}
\caption{Different parameters, dynamical terms and $\dot{P}_{\rm b, int}$ values for PSR J1518$+$4904, the only pulsar with positive $\dot{P}_{\rm b, obs}$, known $\dot{P}_{\rm b, GW}^{Q}$, known proper motion value but unknown parallax. The best distance quoted in the ATNF pulsar catalogue is used here, which is the YMW16 model distance. We also obtain values using the NE2001 model distance. See caption of Table \ref{tb:positivePbdot1bJ1302} for description of columns We do not report the uncertainties here, only the mean values of the simulations are reported.}
\begin{tabular}{l@{\hskip1pt} @{\hskip1pt}l l l@{\hskip1pt} @{\hskip1pt}l@{\hskip2pt} @{\hskip2pt}l@{\hskip2pt} @{\hskip1pt}l@{\hskip1pt} @{\hskip1pt}l}
\hline \hline
 Pulsar & $P_{\rm b}$ & $\mu_T$ & $\dot{P}_{\rm b, obs}$ & $\dot{P}_{\rm b, Gal}$ & $\dot{P}_{\rm b, Shk}$ &  $\dot{P}_{\rm b, int}$ & References  \\ 
 & (days) & (mas/yr) & (${\rm s \, s^{-1}}$) & (${\rm s \, s^{-1}}$) & (${\rm s \, s^{-1}}$) &  (${\rm s \, s^{-1}}$) &  \\ \hline
J1518$+$4904 & 8.6340 & 8.56 & $2.4\times10^{-13}$ & &  &  & \citet{jsk08}\\ \\ 
for $d= 0.964$ kpc &  & &  &  $-1.36\times10^{-13}$ & $1.28\times10^{-13}$ & $2.5\times10^{-13}$   &  \\
(YMW16 model) &  &  & &  &  &  &  \\ \\
    
for $d= 0.631$ kpc &  & &  &  $-1.14\times10^{-13}$ & $8.37\times10^{-14}$ & $2.7\times10^{-13}$   &  \\  
(NE2001 model) &  &  & &  &  &  &  \\  \\ \hline \hline
\label{tb:positivePbdot3a}
\end{tabular}
\end{table*}

\vskip 0.1cm

\textbf{PSR J1518$+$4904}: PSR J1518$+$4904 is a double neutron star system that has $\dot{P}_{\rm b, obs} = 2.4(22) \times 10^{-13}~{\rm s s^{-1}}$. Using the NE2001 model based distance, 0.63(9) kpc, we get $\dot{P}_{\rm b, Gal} = -1.14(6) \times 10^{-13}~{\rm s s^{-1}}$ and $\dot{P}_{\rm b, Shk} = 8.4(12) \times10^{-14}~{\rm s s^{-1}}$ giving $\dot{P}_{\rm b, int} = 2.7(22) \times 10^{-13}~{\rm s s^{-1}}$, and using YMW16 model based distance, 0.96(27) kpc, we get $\dot{P}_{\rm b, Gal} = -1.36(13) \times 10^{-13}~{\rm s s^{-1}} $ and $\dot{P}_{\rm b, Shk} = 1.28(36) \times10^{-13}~{\rm s s^{-1}}$ giving $\dot{P}_{\rm b, int} = 2.5(22) \times 10^{-13}~{\rm s s^{-1}}$. However, as \citet{jsk08} commented that $\dot{P}_{\rm b, obs}$ is only an upper limit, so it is likely that continuing timing analysis will lower the value of $\dot{P}_{\rm b, obs}$ resulting a negative $\dot{P}_{\rm b, int}$. It is also possible that both of the electron density model give wrong values for the distance in the direction of this pulsar. Alternatively, we could also use the expression $\dot{P}_{\rm b, obs} - \dot{P}_{\rm b, Gal} - \dot{P}_{\rm b, GW}^{Q} = \dot{P}_{\rm b, Shk}$ and the definition of $\dot{P}_{\rm b, Shk}$ to place a bound on the distance value $d$. From \citet{jsk08}, using the values of the proper motion $\mu_T = 8.55$ mas/yr, $ \dot{P}_{\rm b, GW}^{Q} = -1.2\times10^{-15}~{\rm s s^{-1}}$, and $\dot{P}_{\rm b, Gal} = -9.5\times10^{-14}~{\rm s s^{-1}}$, we find that $d \leq 0.74$ kpc, which disagrees with the YMW16 model based distance. 

Note that, although the mean value of $\dot{P}_{\rm b, int}$ (for both the electron density models) is larger than $\dot{P}_{\rm b, obs}$, if we consider the uncertainties, it is still possible to have $\dot{P}_{\rm b, int} < \dot{P}_{\rm b, obs}$.

\begin{table*}
\caption{Different parameters, dynamical terms and $\dot{P}_{\rm b, int}$ values for PSR J2129$-$5721, the only pulsar with positive $\dot{P}_{\rm b, obs}$, insufficient PK parameters to calculate $\dot{P}_{\rm b, GW}^{Q}$, known proper motion but unknown parallax. We have obtained the parameters' values using the YMW16 model distance, as well as the NE2001 model distance for the following pulsar. See caption of Table \ref{tb:positivePbdot3a} for description of columns. We do not report the uncertainties here, only the mean values of the simulations are reported.}
\begin{tabular}{l@{\hskip1pt} @{\hskip1pt}l@{\hskip2pt} @{\hskip2pt}l@{\hskip1pt} @{\hskip1pt}l@{\hskip1pt} @{\hskip1pt}l@{\hskip2pt} @{\hskip2pt}l@{\hskip2pt} @{\hskip1pt}l@{\hskip1pt} @{\hskip1pt}l}
\hline \hline
 Pulsar & $P_{\rm b}$ & $\mu_T$ & $\dot{P}_{\rm b, obs}$ & $\dot{P}_{\rm b, Gal}$ & $\dot{P}_{\rm b, Shk}$ &  $\dot{P}_{\rm b, int}$ & References  \\ 
 & (days) & (mas/yr) & (${\rm s \, s^{-1}}$) & (${\rm s \, s^{-1}}$) & (${\rm s \, s^{-1}}$) &  (${\rm s \, s^{-1}}$) &  \\ \hline

J2129$-$5721 & 6.6255 & 13.32 & $7.9\times10^{-13}$ & &  &   & \citet{rhc16} \\ \\

for $d= 6.172$ kpc & & &  &  $-2.56\times10^{-13}$ & $1.52\times10^{-12}$ & $-4.8\times10^{-13}$ &  \\
 (YMW16 model) &  & & &  &  &  &  \\ \\

for $d= 1.363$ kpc & & &  & $-9.29\times10^{-14}$ & $3.36\times10^{-13}$ & $5.5\times10^{-13}$&  \\ 
 (NE2001 model) &  & & &  &  &  &  \\  \\
 \hline \hline
\label{tb:positivePbdot3b}
\end{tabular}
\end{table*}

\vskip 0.1cm

\textbf{PSR J2129$-$5721}: PSR J2129$-$5721 has $\dot{P}_{\rm b, obs} = 7.9(36)\times10^{-13} ~{\rm s s^{-1}}$. Using the NE2001 model based distance, 1.36(51) kpc, we get $\dot{P}_{\rm b, Gal} = -9.3(24) \times 10^{-14}~{\rm s s^{-1}}$ and $\dot{P}_{\rm b, Shk} = 3.4(13) \times10^{-13}~{\rm s s^{-1}}$ giving $\dot{P}_{\rm b, int} = 5.5(38) \times 10^{-13}~{\rm s s^{-1}}$, and using YMW16 model based distance, 6(19) kpc, we get $\dot{P}_{\rm b, Gal} = -2.6(14) \times 10^{-13}~{\rm s s^{-1}} $ and $\dot{P}_{\rm b, Shk} = 1.5(46) \times10^{-12}~{\rm s s^{-1}}$ giving $\dot{P}_{\rm b, int} = -5(47) \times 10^{-13}~{\rm s s^{-1}}$. PSR J2129$-$5721 does not have any parallax measurement and also lacks in sufficient number of measured PK parameters. As a result $\dot{P}_{\rm b, GW}^{Q}$ cannot be obtained for this pulsar.

\citet{rhc16}, however, estimated distance from $\dot{P}_{\rm b, Shk} = \dot{P}_{\rm b, obs} - \dot{P}_{\rm b, Gal}$. But, as we can see, this estimate does not take into account $\dot{P}_{\rm b, GW}^{Q}$. Hence, even as \citet{rhc16} mentions, this is not a good estimate.

\section{Correcting $\dot{P}_{\rm s, obs}$ and looking for problematic $\dot{P}_{\rm s, int}$}

Now, we investigate potential interesting cases where $\dot{P}_{\rm s, obs} $ might be significantly different than $\dot{P}_{\rm s, int} $. $\dot{P}_{\rm s, int} $ is expected to be positive for a rotation powered radio pulsar, as its electromagnetic emission comes at the cost of the rotational kinetic energy. We look for the cases where the ATNF pulsar catalogue and the public data of the PTAs give negative $\dot{P}_{\rm s, obs} $ values and the cases where even though $\dot{P}_{\rm s, obs} $ is positive, $\dot{P}_{\rm s, int} $ turns out to be negative.

\subsection{Negative $\dot{P}_{\rm s, obs}$}

We find four pulsars with anomalous observed values, i.e., negative $\dot{P}_{\rm s, obs} $ in the version $1.60$ of the ATNF pulsar catalogue. There were no additional pulsar with negative $\dot{P}_{\rm s, obs} $ in the public data of PTAs. As $\dot{P}_{\rm s, Shk}$ is always positive, only a significantly negative value of $\dot{P}_{\rm s, Gal}$ can give a positive $\dot{P}_{\rm s, int}$ when $\dot{P}_{\rm s, obs}$ is negative. Unfortunately, among the four pulsars with negative $\dot{P}_{\rm s, obs} $, i.e., PSR J1801$-$3210, PSR J1144$-$6146, PSR J1817$-$0743, and PSR J1829$-$1011, the proper motion of only the first one is known, that too with large uncertainties. So we exclude the last three pulsars from our discussion. Also, for PSR J1801$-$3210, there is no parallax measurement, so we have to use dispersion measure based distance values.

In Table \ref{tb:negativePsdot3b}, we report various dynamical terms calculated for PSR J1801$-$3210 using GalDynPsr and the dispersion measure based distance estimates from both YMW16 and NE2001 models of the Galactic electron density. Using the NE2001 model based distance, 4.03(83) kpc, we get $\dot{P}_{\rm s, Gal} = 4.36(127) \times 10^{-21}~{\rm s s^{-1}}$ and $\dot{P}_{\rm s, Shk} = 1.35(164) \times10^{-20}~{\rm s s^{-1}}$ giving $\dot{P}_{\rm s, int} = -1.8(17) \times 10^{-20}~{\rm s s^{-1}}$, and using YMW16 model based distance 6.1(36) kpc we get $\dot{P}_{\rm s, Gal} = -1.83(871) \times 10^{-21}~{\rm s s^{-1}}$ and $\dot{P}_{\rm s, Shk} = 2.05(274) \times10^{-20}~{\rm s s^{-1}}$ giving $\dot{P}_{\rm s, int} = -1.9(29) \times 10^{-20}~{\rm s s^{-1}}$.

\citet{nbb14} also reported a negative $\dot{P}_{\rm s, int} = -2.7(17)\times10^{-20}~{\rm s \, s^{-1}}$ for this pulsar. They commented that there could be a net radial acceleration towards the Earth which can contribute to the negative $\dot{P}_{\rm s, int}$ of this pulsar. They also hypothesized possible reasons behind such an accelerations, e.g., the presence of near-by massive enough stars, the proximity of any giant molecular cloud (GMCs), or the presence of an orbiting object making PSR J1801$-$3210 a triple system.   

We would like to re-investigate these three pulsars if in the future proper motions and parallaxes are measured.

\begin{table*}
\caption{Different parameters, dynamical terms and $\dot{P}_{\rm s, int}$ value for PSR J1801$-$3210, the only pulsar with negative $\dot{P}_{\rm s, obs}$, known proper motion and unmeasured parallax. The best distance quoted in the ATNF pulsar catalogue is used here, which is the YMW16 model distance value. We also obtain values using the NE2001 model distance. The columns indicate the pulsar name, the spin period ($P_{\rm s}$), the total proper motion ($\mu_T$), the observed rate of change of the spin period ($\dot{P}_{\rm s, obs}$), the rate of change of the spin period caused by the gravitational field of the Galaxy ($\dot{P}_{\rm s, Gal}$), the proper motion contribution to the rate of change of the spin period ($\dot{P}_{\rm s, Shk}$), the calculated intrinsic rate of change of spin period derivative ($\dot{P}_{\rm s, int}$), and the references. We do not report the uncertainties here, only the mean values of the simulations are reported.}
\begin{tabular}{l@{\hskip2pt} @{\hskip2pt}l@{\hskip2pt} @{\hskip2pt}l@{\hskip2pt} @{\hskip2pt}l@{\hskip2pt} @{\hskip2pt}l@{\hskip2pt} @{\hskip2pt}l@{\hskip2pt} @{\hskip2pt}l@{\hskip2pt} @{\hskip2pt}l}

\hline \hline
 Pulsar & $P_{\rm s}$ & $\mu_T$ & $\dot{P}_{\rm s, obs}$ & $\dot{P}_{\rm s, Gal}$ & $\dot{P}_{\rm s, Shk}$ &  $\dot{P}_{\rm s, int}$ & References  \\ 
 & (ms) & (mas/yr) & (${\rm s \, s^{-1}}$) & (${\rm s \, s^{-1}}$) & (${\rm s \, s^{-1}}$) &  (${\rm s \, s^{-1}}$) &  \\ \hline
J1801$-$3210 & 7.4536 & 14 & $-4.4\times 10^{-23}$ & &  &  & \citet{nbb14} \\ \\
for $d= 6.112$ kpc &  &  & & $-1.83\times10^{-21}$ & $2.05\times10^{-20}$ & $-1.87\times10^{-20}$ &  \\ 
(YMW16 model) &  &  &  & & &  &  \\  \\
 
for $d= 4.026$ kpc &  & & & $4.36\times10^{-21}$ & $1.35\times10^{-20}$ & $-1.79\times10^{-20}$ &  \\ 
(NE2001 model) &  &  &  & & &  &  \\ \\

 \hline \hline
\label{tb:negativePsdot3b}
\end{tabular}
\end{table*}

\subsection{Positive $\dot{P}_{\rm s, obs}$ but negative $\dot{P}_{\rm s, int}$}

Finally, we calculate $\dot{P}_{\rm s, int}$ for all pulsars with reported positive values of $\dot{P}_{\rm s, obs}$ and check whether for any of these, $\dot{P}_{\rm s, int}$ turns out to be negative. We find 10 such pulsars from the ones listed in the ATNF pulsar catalogue, three of them form part of PTA data too. %8

For one such pulsar, PSR J1832-0836, if we use the best distance quoted in the ATNF pulsar catalogue (2.50 kpc) for our calculations, we do get a negative $\dot{P}_{\rm s, int}$. However, this pulsar is a NANOGrav pulsar. So, we used the NANOGrav timing solution for this pulsar to obtain LK-bias corrected distance (2.146 kpc) and consequently, obtained a positive $\dot{P}_{\rm s, int}~( 5.55\times 10^{-22}~{\rm s \, s^{-1}})$. Since, there is no anomaly of sign here, we have removed this pulsar from our discussion. We also exclude redback PSR J1622$-$0315 \citep{sa16} and black-widow PSR J1641$+$8049 \citep{rl18}, where the gravitational pull from the ablated material might affect the rate of change of the spin period.

After removing the above mentioned pulsars, we consider the remaining seven pulsars as the actual problematic cases. Among these pulsars, only one has reported parallax and proper motion values (PSR J1024$-$0719). Various dynamical terms for this pulsar are reported in Table \ref{tb:pos_obs_neg_int1a}. The remaining 6 do not have any reported any parallax measurement. The values of various dynamical terms as well as $\dot{P}_{\rm s, int}$ are reported in Table \ref{tb:pos_obs_neg_int3b}. We calculate $\dot{P}_{\rm s, int}$ for the distances obtained from both, the YMW16 model, and the NE2001 model.   

\vskip 0.1cm

\textbf{PSR J1024$-$0719}$^*$: Using $d_{\Pi,\rm LK} = 1.083(226)$ kpc and $\dot{P}_{\rm s, obs} = 1.8553(4)\times10^{-20}~{\rm s s^{-1}}$, as reported by \citet{dcl16}, we obtain $\dot{P}_{\rm s, Gal} = -7.02(37)\times10^{-2}~{\rm s s^{-1}}$ and $\dot{P}_{\rm s, Shk} = 4.8(10)\times 10^{-20}~{\rm s s^{-1}}$ giving $ \dot{P}_{\rm s, int} = -2.9(10)\times 10^{-20}~{\rm s s^{-1}}$. 

For this pulsar, \citet{dcl16} also reported a negative $\dot{P}_{\rm s, int} = -2.9\times 10^{-20}~{\rm s \, s^{-1}}$ which matches with our result in Table \ref{tb:pos_obs_neg_int1a}. They attributed this negative result to a possible presence of a near-by star in a wide orbit with PSR J1024$-$0719. This star may be causing a relative acceleration (calculated by \citet{dcl16} to be $1.7\times10^{-9}~{\rm m s^{-2}}$) along the line of sight, which may be the causing the negative $\dot{P}_{\rm s, int}$. \citet{kp16} also argued that PSR J1024$-$0719 is in a long period (2-20 kyr) binary with a low mass main sequence star. 

Using the recent distance measurement of this pulsar ($1.272$ kpc) based on the \textit{Gaia} DR2 \citep{ming18} and the rest of the parameters from the EPTA timing solution given in \citet{dcl16}, we get $\dot{P}_{\rm s, int} = -3.76\times10^{-20}~{\rm s s^{-1}}$.

\begin{table*}
\caption{Different parameters, dynamical terms and $\dot{P}_{\rm s, int}$ value for PSR J1024$-$0719, the only pulsar with positive $\dot{P}_{\rm s, obs}$ but negative $\dot{P}_{\rm s, int}$, and both, the parallax and the proper motion known. See caption of Table \ref{tb:negativePsdot3b} for description of columns. We do not report the uncertainties here, only the mean values of the simulations are reported.}
\begin{tabular}{l@{\hskip2pt} @{\hskip3pt}l@{\hskip3pt} @{\hskip3pt}l@{\hskip3pt} @{\hskip3pt}l@{\hskip3pt} @{\hskip3pt}l@{\hskip3pt} @{\hskip3pt}l@{\hskip3pt} @{\hskip3pt}l@{\hskip3pt} @{\hskip3pt}l}
\hline \hline
 Pulsar & $P_{\rm s}$ & $\mu_T$ & $\dot{P}_{\rm s, obs}$ & $\dot{P}_{\rm s, Gal}$ & $\dot{P}_{\rm s, Shk}$ &  $\dot{P}_{\rm s, int}$ & References  \\ 
 & (ms) & (mas/yr) & (${\rm s \, s^{-1}}$) & (${\rm s \, s^{-1}}$) & (${\rm s \, s^{-1}}$) &  (${\rm s \, s^{-1}}$) &  \\ \hline
J1024$-$0719 & 5.1622 & 59.72 & $1.86\times 10^{-20}$ & $-7.02\times 10^{-22}$ & $4.84\times 10^{-20}$ & $-2.92\times 10^{-20}$ & \citet{dcl16} \\
(for $d = 1.083$ kpc) & & & & & & &  \\  \\
 \hline \hline
\label{tb:pos_obs_neg_int1a}
\end{tabular}
\end{table*}

\vskip 0.1cm

\textbf{PSR J1142$+$0119}: \citet{sa16} reports PSR J1142$+$0119 to have $\dot{P}_{\rm s, obs} = 14.99\times10^{-21} ~{\rm s s^{-1}}$. Using the NE2001 model based distance, 1.7(292) kpc, we get $\dot{P}_{\rm s, Gal} = -8.8(8)\times 10^{-22}~{\rm s s^{-1}}$ and $\dot{P}_{\rm s, Shk} = 1.2(14) \times10^{-19}~{\rm s s^{-1}}$ giving $\dot{P}_{\rm s, int} = -1(1) \times 10^{-19}~{\rm s s^{-1}}$, and using YMW16 model based distance 2.17(242) kpc we get $\dot{P}_{\rm s, Gal} = -8.1(75) \times 10^{-22}~{\rm s s^{-1}}$ and $\dot{P}_{\rm s, Shk} = 3.0(48) \times10^{-19}~{\rm s s^{-1}}$ giving $\dot{P}_{\rm s, int} = -2.8(48) \times 10^{-19}~{\rm s s^{-1}}$.

We have used the timing solutions, including DM and proper motion, provided in \citet{sa16} for the above calculations. \citet{sa16} also reports $\dot{P}_{\rm s, int} = 14.5(1) \times 10^{-21}~{\rm s s^{-1}}$. However, they don't include $\dot{P}_{\rm s, Shk}$ and calculate as $\dot{P}_{\rm s, int} = \dot{P}_{\rm s, obs} - \dot{P}_{\rm s, Gal}$. They attribute this exclusion to $\approx2\sigma$ uncertainty in proper motion values.

\vskip 0.1cm

\textbf{PSR J1327$-$0755}: 
PSR J1327$-$0755 has $\dot{P}_{\rm s, obs} = 1.77\times10^{-21} ~{\rm s s^{-1}}$, as reported in \citet{blr13}. Using the NE2001 model based distance, 0.86(22) kpc, we get $\dot{P}_{\rm s, Gal} = -7.6(48) \times 10^{-22}~{\rm s s^{-1}}$ and $\dot{P}_{\rm s, Shk} = 1(19) \times10^{-19}~{\rm s s^{-1}}$ giving $\dot{P}_{\rm s, int} = -9(186) \times 10^{-20}~{\rm s s^{-1}}$, and using YMW16 model based distance 25(25) kpc we get $\dot{P}_{\rm s, Gal} = -9.8(41) \times 10^{-21}~{\rm s s^{-1}}$ and $\dot{P}_{\rm s, Shk} = 1.6(22) \times10^{-18}~{\rm s s^{-1}}$ giving $\dot{P}_{\rm s, int} = -1.6(22) \times 10^{-18}~{\rm s s^{-1}}$.

 \citet{blr13} reported for this pulsar a large composite proper motion ($99 \pm 23 ~{\rm mas yr^{-1}}$) and claimed that given this large value either the distance estimate by NE2001 model is inaccurate or additional observational parameters are not taken into account.

\vskip 0.1cm

\textbf{PSR J1405$-$4656}: 
PSR J1405$-$4656 is reported to have $\dot{P}_{\rm s, obs} = 2.79\times10^{-20} ~{\rm s s^{-1}}$ \citep{btb15}. Using the NE2001 model based distance, 0.58(7) kpc, we get $\dot{P}_{\rm s, Gal} = -1.47(19) \times 10^{-22}~{\rm s s^{-1}}$ and $\dot{P}_{\rm s, Shk} = 2.50(78) \times10^{-20}~{\rm s s^{-1}}$ giving $\dot{P}_{\rm s, int} = 3.0(78) \times 10^{-21}~{\rm s s^{-1}}$, and using YMW16 model based distance, 0.67(9) kpc we get $\dot{P}_{\rm s, Gal} = -1.74(23) \times 10^{-22}~{\rm s s^{-1}} $ and $\dot{P}_{\rm s, Shk} = 2.89(90) \times10^{-20}~{\rm s s^{-1}}$ giving $\dot{P}_{\rm s, int} = -7.9(905) \times 10^{-22}~{\rm s s^{-1}}$.

\citet{btb15} reported that the error in $\dot{P}_{\rm s, Shk}$ calculated by them (and consequently in $\dot{P}_{\rm s, int}$) is large due to errors in the proper motion values and in the distance derived from the electron density model. 

\vskip 0.1cm

\textbf{PSR J1721$-$2457}: 
PSR J1721$-$2457 is reported to have $\dot{P}_{\rm s, obs} = 5.56\times10^{-21} ~{\rm s s^{-1}}$ \citep{dcl16}. Using the NE2001 model based distance, 1.30(16) kpc, we get $\dot{P}_{\rm s, Gal} = 4.58(74) \times 10^{-22}~{\rm s s^{-1}}$ and $\dot{P}_{\rm s, Shk} = 6.9(89) \times10^{-21}~{\rm s s^{-1}}$ giving $\dot{P}_{\rm s, int} = -1.8(89)\times 10^{-21}~{\rm s s^{-1}}$, and using YMW16 model based distance, 1.40(63) kpc, we get $\dot{P}_{\rm s, Gal} = 5.1(29) \times 10^{-22}~{\rm s s^{-1}} $ and $\dot{P}_{\rm s, Shk} = 7.4(100) \times10^{-21}~{\rm s s^{-1}}$ giving $\dot{P}_{\rm s, int} = -2.4(100) \times 10^{-21}~{\rm s s^{-1}}$.

 \citet{dcl16}, however, gave the value of $\dot{P}_{\rm s, int} = 0.0(7)\times10^{-20}~{\rm s \, s^{-1}}$ in it's Table 6.
 
\vskip 0.1cm 

\textbf{PSR J1813$-$2621}: 
\citet{lem15} reports that PSR J1813$-$2621 has $\dot{P}_{\rm s, obs} = 1.25\times10^{-20} ~{\rm s s^{-1}}$. Using the NE2001 model based distance, 2.71(47) kpc, we get $\dot{P}_{\rm s, Gal} = 1.53(37) \times 10^{-21}~{\rm s s^{-1}}$ and $\dot{P}_{\rm s, Shk} = 1.6(21) \times10^{-20}~{\rm s s^{-1}}$ giving $\dot{P}_{\rm s, int} = -4.7(207) \times 10^{-21}~{\rm s s^{-1}}$, and using YMW16 model based distance, 3.01(39) kpc, we get $\dot{P}_{\rm s, Gal} = -1.76(32) \times 10^{-21}~{\rm s s^{-1}} $ and $\dot{P}_{\rm s, Shk} = 1.7(23) \times10^{-20}~{\rm s s^{-1}}$ giving $\dot{P}_{\rm s, int} = -6.7(229) \times 10^{-21}~{\rm s s^{-1}}$.

For this pulsar, \citet{lem15} gave a negative value of $\dot{P}_{\rm s, int}$, i.e., $\dot{P}_{\rm s, int} = -0.6(17)\times10^{-21}~{\rm s \, s^{-1}}$. But they mentioned that they were unable to measure a significant proper motion in declination because of the low ecliptic latitude of the pulsar.

\vskip 0.1cm

\textbf{PSR J1843$-$1448}: 
PSR J1843$-$1448 has $\dot{P}_{\rm s, obs} = 6.21\times10^{-21} ~{\rm s s^{-1}}$, as reported in \citet{lem15}. Using the NE2001 based distance, 2.99(53) kpc, we get $\dot{P}_{\rm s, Gal} = 1.34(25) \times 10^{-21}~{\rm s s^{-1}}$ and $\dot{P}_{\rm s, Shk} = 1.0(15) \times10^{-20}~{\rm s s^{-1}}$ giving $\dot{P}_{\rm s, int} = -5.3(145) \times 10^{-21}~{\rm s s^{-1}}$, and using YMW16 model based distance, 3.47(69) kpc, we get $\dot{P}_{\rm s, Gal} = 1.47(26) \times 10^{-21}~{\rm s s^{-1}} $ and $\dot{P}_{\rm s, Shk} = 1.2(17) \times10^{-20}~{\rm s s^{-1}}$ giving $\dot{P}_{\rm s, int} = -7(17) \times 10^{-21}~{\rm s s^{-1}}$.

For this pulsar, \citet{lem15} gave the negative values of $\dot{P}_{\rm s, int}$, i.e., $\dot{P}_{\rm s, int} = -0.5(11)\times10^{-21}~{\rm s \, s^{-1}}$. But like PSR J1813$-$2621, here too, they mentioned that they were unable to measure a significant proper motion in declination because of the low ecliptic latitude of the pulsar.\\

We know the fact that $\dot{P}_{\rm s, int}$ can not be negative. So all of the above cases where we found a negative value for $\dot{P}_{\rm s, int}$, mean either the measurements of proper motion and/or the parallax is erroneous, or there exist unknown additional sources of significantly large negative acceleration of the pulsars (as discussed for the case of PSR J1801$-$3210). Further study exploring this issue will be interesting.

\begin{table*}
\caption{Different parameters, dynamical terms and $\dot{P}_{\rm s, int}$ values for the pulsars with positive $\dot{P}_{\rm s, obs}$ but negative $\dot{P}_{\rm s, int}$. These pulsars have known proper motions but unknown parallaxes. The best distance quoted in the ATNF pulsar catalogue is used here, which is the YMW16 model distance. We also obtain values using the NE2001 model distance. See caption of Table \ref{tb:negativePsdot3b} for description of columns. We do not report the uncertainties here, only the mean values of the simulations are reported.}
\begin{tabular}{l@{\hskip2pt} @{\hskip3pt}l@{\hskip3pt} @{\hskip3pt}l@{\hskip3pt} @{\hskip3pt}l@{\hskip3pt} @{\hskip3pt}l@{\hskip3pt} @{\hskip3pt}l@{\hskip3pt} @{\hskip3pt}l@{\hskip3pt} @{\hskip3pt}l}
\hline \hline
 Pulsar & $P_{\rm ms}$ & $\mu_T$ & $\dot{P}_{\rm s, obs}$ & $\dot{P}_{\rm s, Gal}$ & $\dot{P}_{\rm s, Shk}$ &  $\dot{P}_{\rm s, int}$ & References  \\ 
 & (ms) & (mas/yr) & (${\rm s \, s^{-1}}$) & (${\rm s \, s^{-1}}$) & (${\rm s \, s^{-1}}$) &  (${\rm s \, s^{-1}}$) &  \\ \hline
 
J1142$+$0119 & 5.0753 & 105 & $1.499\times 10^{-20}$ & &  & & \citet{sa16}\\  \\
for $d= 2.17$ kpc &  & &  &  $-8.11\times10^{-22}$ & $2.97\times10^{-19}$ & $-2.81\times10^{-19}$  &  \\
(YMW16 model) &  &  &  & &  &  &  \\ \\ 
   
for $d= 0.86$ kpc &  & & &  $-8.80\times10^{-22}$ & $1.18\times10^{-19}$ & $-1.02\times10^{-19}$ &  \\ 
(NE2001 model) &  &  &  & &  &  &  \\  \\ \hline
 
J1327$-$0755 & 2.6779 & 99 & $1.773\times 10^{-21}$ & &  & & \citet{blr13}\\  \\
for $d= 25.0$ kpc &  & &  &  $-9.84\times10^{-22}$ & $1.59\times10^{-18}$ & $-1.57\times10^{-18}$  &  \\
(YMW16 model) &  &  &  & &  &  &  \\ \\ 
    
for $d= 1.735$ kpc &  & & &  $-7.58\times10^{-22}$ & $1.10\times10^{-19}$ & $-9.16\times10^{-20}$ &  \\ 
(NE2001 model) &  &  &  & &  &  &  \\  \\ \hline

J1405$-$4656 & 7.6022 & 48 & $2.79\times10^{-20}$  &  & & & \citet{btb15}\\ \\
for $d= 0.6691$ kpc &  & &  &  $-1.74\times10^{-22}$ & $2.89\times10^{-20}$ & $-7.91\times10^{-22}$  &  \\ 
(YMW16 model) &  &  &  & & &  &  \\ \\
   
for $d= 0.580$ kpc &  &  & &  $-1.47\times10^{-22}$ & $2.50\times10^{-20}$ & $3.03\times10^{-21}$  &  \\ 
(NE2001 model) &  &  &  & & &  &  \\  \\ \hline

J1721$-$2457 & 3.4966 & 25 & $5.56\times10^{-21}$ & &  & & \citet{dcl16} \\ \\
for $d= 1.398$ kpc &  &  & &  $5.11\times10^{-22}$ & $7.47\times10^{-21}$ & $-2.42\times10^{-21}$  &  \\
(YMW16 model) &  &  &  & & &  &  \\  \\
   
for $d= 1.299$ kpc &  & & &  $4.58\times10^{-22}$ & $6.94\times10^{-21}$ & $-1.83\times10^{-21}$ &  \\ 
(NE2001 model) &  &  & & &  &  &  \\  \\ \hline

J1813$-$2621 & 4.4300 & 23 & $1.2466\times10^{-20}$ & &  & & \citet{lem15} \\ \\
for $d= 3.013$ kpc &  &  & & $-1.76\times10^{-21}$ & $1.74\times10^{-20}$ & $-6.71\times10^{-21}$  &  \\
(YMW16 model) &  &  &  & & &  &  \\ \\
  
for $d= 2.709$ kpc &  &  & &  $1.53\times10^{-21}$ & $1.57\times10^{-20}$ & $-4.73\times10^{-21}$ &  \\  
(NE2001 model) &  &  &  & & &  &  \\  \\ \hline

J1843$-$1448 & 5.4713 & 16 & $6.209\times10^{-21}$ & &  & & \citet{lem15} \\  \\
for $d= 3.472$ kpc  &  & &  & $1.47\times10^{-21}$ & $1.17\times10^{-20}$ & $-6.99\times10^{-21}$  &  \\ 
(YMW16 model) &  &  &  & & &  &  \\ \\
 
for $d= 2.995$ kpc &  &  & &  $1.34\times10^{-21}$ & $1.01\times10^{-20}$ & $-5.25\times10^{-21}$  &  \\ 
(NE2001 model) &  &  &  & &  &  &  \\  \\ 
 \hline \hline
\label{tb:pos_obs_neg_int3b}
\end{tabular}
\end{table*}

\vskip 0.6cm

\vspace{-2em}
\section{Conclusion}

As we know, for rotation powered pulsars, the rate of change of the spin period is expected to be positive. On the other hand, for a clean binary system (where gravity is the only force acting on the pulsar) the rate of change of the orbital period is expected to be negative.

In the present paper, we explored the pulsars with negative $\dot{P}_{\rm s, obs}$ and positive $\dot{P}_{\rm b, obs}$ and checked whether the intrinsic values of these parameters have the correct sign. We have also looked for the pulsars for which although $\dot{P}_{\rm s, obs}$ and $\dot{P}_{\rm b, obs}$ seemed to have the correct signs, $\dot{P}_{\rm s, int}$ and $\dot{P}_{\rm b, int}$ ended up with wrong signs.

We found 17 pulsars with positive $\dot{P}_{\rm b, obs}$ in the ATNF pulsar catalogue (version 1.60), three  pulsars with positive $\dot{P}_{\rm b, obs}$ in the version-2 data release of IPTA, and one additional pulsar with positive $\dot{P}_{\rm b, obs}$ in \citet{zsd15}. There was no case of negative $\dot{P}_{\rm b, obs}$, yet positive $\dot{P}_{\rm b, int}$ values. For the cases of anomalous spin period derivative values, we found four pulsars with negative $\dot{P}_{\rm s, obs}$ in the ATNF pulsar catalogue (version 1.60) and seven cases in all (ATNF and all PTAs' data combined) with a positive $\dot{P}_{\rm s, obs}$, but a negative $\dot{P}_{\rm s, int}$. 

We would like to point out that ATNF had reported a wrong value of $\dot{P}_{\rm s, obs}$ for PSR J2222$-$0137 as $5.8024\times 10^{-21}~{\rm s \, s^{-1}}$ which was giving an anomalous negative $\dot{P}_{\rm s, int}$, whereas, in literature it is $5.8024\times 10^{-20}~{\rm s \, s^{-1}}$ (confirmed from \citet{cfg17} and references therein). This does not give any anomalous result, i.e., it gives a positive value of $\dot{P}_{\rm s, int} = 1.73\times 10^{-20}~{\rm s \, s^{-1}}$, which is very close to the value given by \citet{cfg17} ($1.75\times 10^{-20}~{\rm s \, s^{-1}}$). Hence, we removed J2222$-$0137 from our discussion.

We used Model-La of GalDynPsr for calculating the intrinsic period derivatives. We were able to obtain negative $\dot{P}_{\rm b, int}$ for five of the pulsars with positive $\dot{P}_{\rm b, obs}$ including the case of PSR J2129-5721 where negative $\dot{P}_{\rm b, int}$ is obtained when YMW16 model distance is used. Even for the rest of the cases, with anomalous observed period derivative values, where we could not obtain the correct sign, we were able to reduce the magnitude and shift the values towards the correct sign. An exception for this were PSR J0900$-$3144 and PSR J1518$+$4904 where $\dot{P}_{\rm b, int}$ did not reduce in magnitude to $\dot{P}_{\rm b, obs}$. However, for both of these pulsars, if we consider the uncertainties, it is still possible to have $\dot{P}_{\rm b, int} < \dot{P}_{\rm b, obs}$. Moreover, for PSR J0900$-$3144, even if $\dot{P}_{\rm b, int} > \dot{P}_{\rm b, obs}$, that might be explained due to the proximity of the Gum Nebula, as reported by \citet{pg15}, which could be the cause of extra acceleration  causing the discrepancy.

We calculated $\dot{P}_{\rm b, GW}^{Q}$ for the cases where enough PK parameters were provided and used it to obtain $\dot{P}_{\rm b, Shk}$, and consequently, an alternate distance estimate $d_{\rm Shk}$. We compared these distance values with those obtained from the parallax values (Table \ref{tb:positivePbdot1a}). 

It is important to perform detailed observations around the pulsars with anomalous values of $\dot{P}_{\rm s, int}$, and $\dot{P}_{\rm b, int}$. We need to look for near-by stars, molecular clouds, and probable existence of additional gravitationally bound companions. As these factors, if they exist, can cause the pulsar to accelerate in such a way that it contributes to $\dot{P}_{\rm s, obs}$ and $\dot{P}_{\rm b, obs}$ values. Subtracting the effects of such factors can lead to a much accurate estimation of the intrinsic period derivatives. 

Future technological advancements will lead to detections, and improvements in measurements (if already detected) of parallax, and proper motion values. This will further improve the intrinsic period derivative calculations.

%%%%%%%%%%%%%%%%%%%% REFERENCES %%%%%%%%%%%%%%%%%%

%%%%%%%%%%%%%%%%%%%%%%%%%%%%%%%%%%%%%%%%%%%%%%%%%%

% Don't change these lines
\bsp	% typesetting comment
\label{lastpage}
\end{document}